\documentclass[a4paper]{article}

\usepackage{INTERSPEECH2021}
\usepackage{url}
\usepackage{graphicx} % table scaling
\usepackage{enumitem}

\title{Pathological voice adaptation with autoencoder-based voice conversion}

\name{Marc Illa$^{*1, 2}$, Bence Mark Halpern$^{*1,3,4}$, Rob van Son$^{3,4}$, Laureano Moro-Velázquez$^{5}$, Odette Scharenborg$^1$
\thanks{$^*$Equal contribution.
 }
}

\address{
  $^1$Multimedia Computing Group, Delft University of Technology, Delft, The Netherlands \\
  $^2$Universitat Politecnica de Catalunya, Barcelona, Spain\\
  $^3$University of Amsterdam, Amsterdam, The Netherlands  \\
  $^4$Netherlands Cancer Institute, Amsterdam, The Netherlands \\
  $^5$Johns Hopkins University, Baltimore, USA}
\email{\{m.illa,o.e.scharenborg\}@tudelft.nl, \{b.halpern,r.v.son\}@nki.nl, laureano@jhu.edu}

\begin{document}

\maketitle
\begin{abstract}

In this paper, we propose a new approach to pathological speech synthesis. Instead of using healthy speech as a source, we customise an existing pathological speech sample to a new speaker's voice characteristics. This approach alleviates the evaluation problem one normally has when converting typical speech to pathological speech, as in our approach, the voice conversion (VC) model does not need to be optimised for speech degradation but only for the speaker change. This change in the optimisation ensures that any degradation found in naturalness is due to the conversion process and not due to the model exaggerating characteristics of a speech pathology. To show a proof of concept of this method, we convert dysarthric speech using the UASpeech database and an autoencoder-based VC technique. Subjective evaluation results show reasonable naturalness for high intelligibility dysarthric speakers, though lower intelligibility seems to introduce a marginal degradation in naturalness scores for mid and low intelligibility speakers compared to ground truth. Conversion of speaker characteristics for low and high intelligibility speakers is successful, but not for mid. Whether the differences in the results for the different intelligibility levels is due to the intelligibility levels or due to the speakers needs to be further investigated.
\end{abstract}
\noindent\textbf{Index Terms}: voice conversion, pathological speech, variational autoencoder

\section{Introduction}
\label{sec:introduction}

Data-driven speech synthesis has recently been reaching new heights with the introduction of deep neural networks (DNNs). However, the success of these techniques is subject to high quality data and a large quantity of data, either of which is not available for many applications. Pathological speech synthesis, where the goal is to synthesise natural, but pathologically sounding samples, is such an application. 
Pathological speech synthesis has several motivations, the most notable being the data augmentation for automatic speech recognisers (ASRs), where the goal is to generate more data in order to improve recognition of pathological speech \cite{vachhani2018data, harvillsynthesis, jiao2018simulating}. The second motivation for the development of pathological speech synthesis is that it could assist in informed decision making for the medical conditions at the root of the pathology. For instance, oral cancer surgery results in changes to a speaker's voice. The availability of a synthesis model that can generate how the voice could sound after surgery could help the patients and clinicians to make informed decisions about the surgery and alleviate the stress of the patients \cite{halpern2021evaluation, Halpern2020}.

While there are many speech synthesis techniques for typical speech, not many of these are applicable if we wish to synthesise highly natural pathological speech. Formant \cite{rudzicz2013adjusting} and articulatory synthesis \cite{aryal2016data} are lacking in naturalness compared to DNN-based speech synthesis. Text-to-speech techniques (TTS) lack both linguistic resources (i.e a pronunciation lexicon) and the amount of data needed for these problems. The only promising method to synthesise pathological speech seems to be voice conversion (VC), which only needs a relatively small amount of data, compared to neural TTS.

However, synthesising pathological speech via VC is not without challenges. Existing pathological speech corpora \cite{kim2008dysarthric, rudzicz2012torgo, Halpern2020, middag2012automatic} provide healthy control speakers, but healthy speech recordings from the same pathological speaker are rarely available. This means that a successful pathological voice conversion system needs to learn conversion of both, the voice and pathological characteristics simultaneously, as suggested in previous work \cite{halpern2021evaluation}. However, evaluation of such a setup is difficult. This is because the VC system is directly optimised for speech degradation in terms of the pathology, which would need the listeners (the evaluators of these systems) to be able to rate the success of generating the pathological characteristics and the synthetic/natural aspects of the speech separately. As we will show later in this paper, listeners struggle differentiating between speech severity and synthetic aspects of the speech. This can result in two, counter-intuitive scenarios from the viewpoint of typical VC: (1) a pathological VC system that is not able  to properly capture the characteristics of the pathological speech could still receive better naturalness scores than the reference pathological speech; (2) Conversely, a VC system that is able to mimic the pathology, albeit exaggeratedly, could produce a naturalness score that is a lot lower than that of the reference. 

Therefore, we propose a new approach where instead of using healthy speech as source for the VC, we use dysarthric speech, which is already pathological, and the VC system only has to customise it to a new (healthy/dysarthric) speaker's voice characteristics, i.e by using some representation of the speaker (speaker embedding). This synthesis approach alleviates the problem with naturalness ratings as the dysarthric-to-dysarthric VC is not optimised directly for speech degradation, therefore degradation is only due to the synthetic aspects compared to the source pathological utterance. Our first goal is to assess whether we can convert the voice characteristics of the pathological speakers in this setup in a natural way, while simultaneously assessing how natural real pathological speech is perceived.

In order to perform the VC, an autoencoder-based method will be used \cite{Hsu2017}. Autoencoder-based methods are of special interest in clinical scenarios as they are non-parallel, thus allow for incomplete data collection situations, while also being easier to train than GAN-based methods due to well-defined convergence criteria because they have only a single loss \cite{kaneko2018cyclegan, kaneko2019cyclegan, kaneko2020cyclegan}. In this paper, we use HL-VQ-VAE-3 which is a type of variational autoencoder (VAE) using discrete representations. This hierarchical design has recently shown to give better results for VC \cite{ho2020non} than the original VQ-VAE. Furthermore, by conditioning on speaker labels, the model allows to converting to/from multiple speakers within one single model.

An important additional goal of this work is to investigate whether standard VC techniques can be used for non-standard speech. It is well known from other domains of speech technology such as automatic speech recognition (ASR) that standard ASR systems perform poorly on  atypical speech \cite{feng2021quantifying, adda2005speech,moro2019study,koenecke2020racial, hermann2020dysarthric}, making standard speech technology techniques less accessible to people with atypical speech. Our paper is thus also a preliminary investigation of a VQ-VAE-based VC technique’s performance on converting a pathological source utterance instead of a typical utterance from a non-dysarthric speaker.

To summarise, in this paper we train a dysarthric-to-dysarthric VC system to answer the following research questions: (\textbf{RQ1}) \textit{Can we convert the voice characteristics of a pathological speaker to another pathological speaker of the same severity with reasonable naturalness (where reasonable means comparable to non-parallel VC methods on typical speech)? In other words, is VC technology accessible to people with pathological speech?} (\textbf{RQ2}) \textit{How does (real) pathological speech affect the mean opinion score (MOS)? In other words, what is the maximum attainable naturalness of synthetic pathological speech?}

Section \ref{sec:dataset} will start with the discussion of the used UASpeech dataset and the used VQ-VAE methods for the task, and finally concluded by the experimental design to test the approach. 
The perceptual evaluation results are presented in Section \ref{sec:results}, followed by a discussion of the limitations of the proposed method, and further comments on the accessibility of VC to pathological speakers. Some of the samples are available at \textit{https://pathologicalvc.github.io} .

\begin{figure}[ht]
    \centering
    \includegraphics[width=\columnwidth]{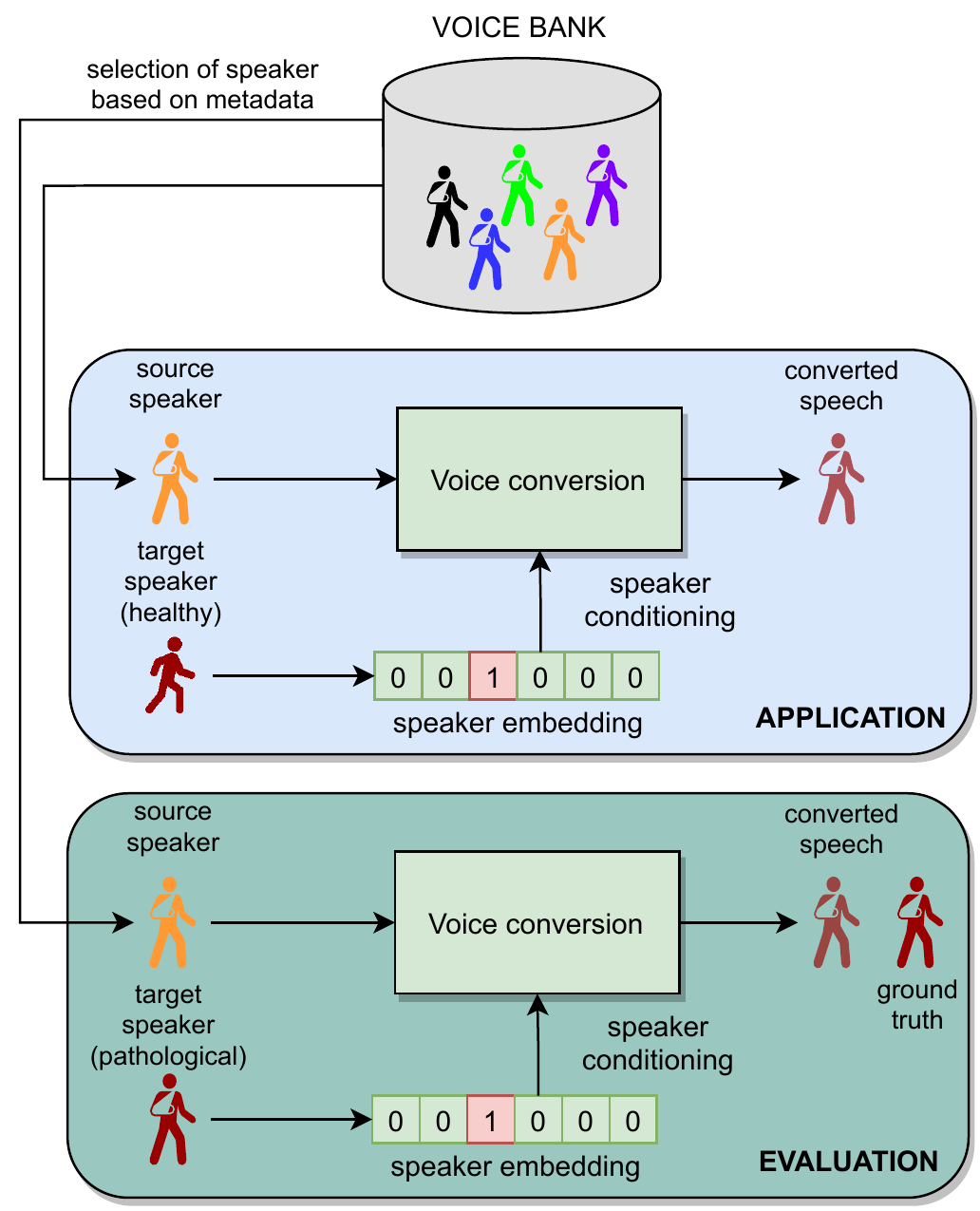}
    \caption{Outline of our approach: the speech from a model pathological speaker is converted into speech with the characteristics of another pathological speaker. Red/orange colours denote the identity of the speaker.  The figure is further explained in Section \ref{sec:dataset}.}
    \label{fig:conversion_setup}
\end{figure}

\section{Design and methods}
\label{sec:dataset}

\subsection{Description of the dataset and preprocessing}

In this study we use the UASpeech corpus \cite{kim2008dysarthric}, which contains isolated-word recordings of 15 speakers with dysarthria. These recordings consist of 449 words which are divided into 3 blocks of equal length (B1, B2 and B3). The speakers are divided into four groups based on their intelligibility: very low, low, mid and high, which correspond to 0-25\%, 25-50\%, 50-75\% and 75-100\% human transcription word error rate (WER) of the recordings, respectively. The transcriptions were done by 5 American English native speakers, who are non-expert listeners.

The vocoder used (see Section \ref{subsec:vcmodel}) is trained using the VCTK dataset \cite{veaux2017cstr}, which contains speech of 108 native English speakers with different accents. The preprocessing consists of downsampling the tracks from 48 kHz to 24 kHz, which is done with librosa \cite{mcfee2015librosa}.

The UASpeech data is preprocessed following \cite{harvillsynthesis}: stationary noise is removed using Noisereduce \cite{tim_sainburg_2019_3243139} and the silence from the beginning and end of the clips is cut. Then, the audio is resampled from 16 kHz to 24 kHz and normalised. Finally, 80-dimension mel-spectrograms (similar to \cite{skerry2018towards}) are extracted from the audio files and used to compute the mel-cepstrum, which serves as input to our model. 

\subsection{Voice conversion model}
\label{subsec:vcmodel}

The model is a 3-stage VQ-VAE. In the first stage, the input $x$ to the model is a mel-cepstrum that goes through the convolutional encoder resulting in a hidden variable $u_{1}$ and a latent variable $z_{1}$. The second stage is identical to the first stage, except instead of $x$, now $u_{1}$ is fed into another convolutional encoder, resulting in $u_{2}$ and $z_{2}$. This is repeated for the third stage, feeding $u_{2}$ to obtain $z_{3}$ and $u_{3}$. 
This successive encoding serves to model the features in the speech that are present on successively longer temporal scales. 

The variables $z_{n}$ are all quantised using a nearest neighbour classifier with respect to the codebook's codewords of the corresponding stage. Then, we perform the decoding of the quantised variables $q_{n}$ at each stage. The decoder is also convolutional which is additionally conditioned on a speaker label. During training, a speaker embedding table is learned from the training speakers, and during conversion/inference, this embedding will correspond to the target speaker of the conversion, which we can get by a table lookup. The decoding starts at the third stage and goes back to the first stage. The input of the third stage decoder is $q_{3}$ while for the second and first level the $q_{n}$ signal is concatenated with the output $v_{n}$ of the previous stage (the output $v_{2}$ of the 3rd stage is fed to the 2nd and the output $v_{1}$ of the 2nd is fed to the 1st).

For the conversion, the trained model receives the input mel-cepstrum from a source speaker which is encoded and quantised in the same way as it is during training. Then, the speaker embedding is used to condition the decoder on a target speaker, so the source speaker quantised latent variables $q_{n}$ are decoded conditioned on the target speaker embedding, which results in the decoded mel-cepstrum. Finally, the mel-cepstrum is resynthesised to the speech waveform using a Parallel WaveGAN vocoder\footnote{https://github.com/kan-bayashi/ParallelWaveGAN} \cite{yamamoto2020parallel}.

\subsection{Details of the experimental design}

As a reminder, in this study, we customise pathological speech to a different pathological speaker's voice characteristics. However, the clinical application would need customisation to a healthy speaker's characteristics. In the top panel of Figure \ref{fig:conversion_setup}, the application scenario is visualised, i.e., how the system could be used in a clinical setting. In the bottom part, our proposed evaluation scenario - the experiments that we do in the paper - is illustrated. 

Looking at the top panel, a source pathological speaker is first selected from a large voice bank consisting of many samples of pathological speakers. Based on metadata, a clinical team could decide the kind of pathological speech degradation which is most likely for a patient. In this paper, we pair up by severity, but in actual practice an appropriate source speaker could be found matched by age, region, and type of treatment. This leads to a selection of a source pathological speaker. Using a small amount of a new patient’s voice (target speaker), a speaker embedding can be extracted using the VQ-VAE based technique. Finally, we obtain the converted speech, which is expected to be pathological, but with the new patient’s voice characteristics. The problem is that for the UASpeech, we don't have parallel pre-pathology and post-pathology voices. Therefore, a separate evaluation scheme has to be setup where we assume that the pathological and the healthy speaker embeddings should be unchanged for the same speaker, which is not always true, we refer to further discussion about this in Section \ref{subsec:limitations}.

The evaluation scheme is explained in the bottom panel. To circumvent the problem with the pre-pathology and post-pathology, we change the conversion process for the evaluation as follows. Instead of a new healthy speaker, we enroll a new dysarthric speaker with a matched intelligibility of the speech pathology from the UASpeech dataset because a ground truth (GT) is available there. The converted speech can then be compared to this GT to provide a proof of concept for the system. 

\begin{table}[ht]
\centering
\caption{Speaker pairs used for the VC experiments and their subjective WER differences.}
\resizebox{\columnwidth}{!}{%
\begin{tabular}{|c|c|c|}
\hline
\textbf{Speaker A (WER\%)} & \textbf{Speaker B (WER\%)} & \textbf{$\Delta$WER (\%)} \\ \hline
M04 (2\%) & M12 (7.4\%) & 5.4\% \\ \hline
M05 (58\%) & M11 (62\%) & 4\% \\ \hline
M08 (93\%) & M10 (93\%) & 0\% \\ \hline
\end{tabular}
}
\label{table:speaker_pairs}

\end{table}

In our experiments, we convert the speech of three speaker pairs in both directions
The setup for the experiments is the following. We train the VC model with all B1 and B3 sets of words of every dysarthric speaker to stay consistent with the standard UASpeech train-test partitioning. 

We perform VC on the speech from B2 between speakers with a similar level of dysarthria. The selected dysarthric speaker pairs along with their corresponding human transcription error rates from UASpeech are summarised in Table \ref{table:speaker_pairs}. Unfortunately, it has not been possible to include females speakers because all female speakers had a different severity in the UASpeech dataset. %Other important factors such as the type of dysarthria are not controlled in this design. 
We also refrained from controlling for the type of dysarthria in our experimental design, as that would have led to certain speaker pairs having excessive difference in their intelligibility, which would contrive the aim of the paper.

\subsection{Subjective evaluation experiments}

In order to answer our research questions, we performed subjective evaluation experiments. For RQ1 a subjective speaker similarity experiment was carried out, while for RQ2 a subjective naturalness experiment was carried out. The design of these experiments (including the composition of different stimuli) closely follow those of the VCC challenge standards \cite{yi2020voice, toda2016voice}. These experiments were run on the Qualtrics platform, and the participants (10 American English native listeners) were recruited through Prolific. All participants were remunerated justly (7.80 GBP per hour).

For the naturalness experiment, we used a mean opinion score (MOS) naturalness test. We hypothesised that listeners will not be able to distinguish between the distortions in the audio and the pathological characteristics of the speech. In order to account for this, we included GT stimuli in the naturalness test, which allows direct comparison of naturalness with real samples. The GT shows the maximum attainable naturalness (second part of RQ2) and the differences of the GT and VC scores show the reduction due to the synthetic aspects. To answer the first part of RQ2, we included healthy, natural stimuli, which allows us to measure the reduction in naturalness due to the reduction intelligibility. Nevertheless, we encouraged listeners to ignore the atypical aspects of the speech by adopting the naturalness question from the VCC2020 \cite{yi2020voice}, which was proposed for cross-lingual VC, where pronunciation errors could appear, similar to pathological speech. For the speaker similarity test, we used an AB test in which listeners were asked to listen to two stimuli, and indicate if they thought they came from the same speaker, and rate their confidence in this decision. The question for the speaker similarity was directly adopted from the VCC2016 challenge \cite{toda2016voice}.

\section{Results and discussion}
\label{sec:results}

\subsection{Naturalness}
\label{sec:naturalness_result}
\begin{figure}[ht]
    \centering
    \includegraphics[width=\columnwidth]{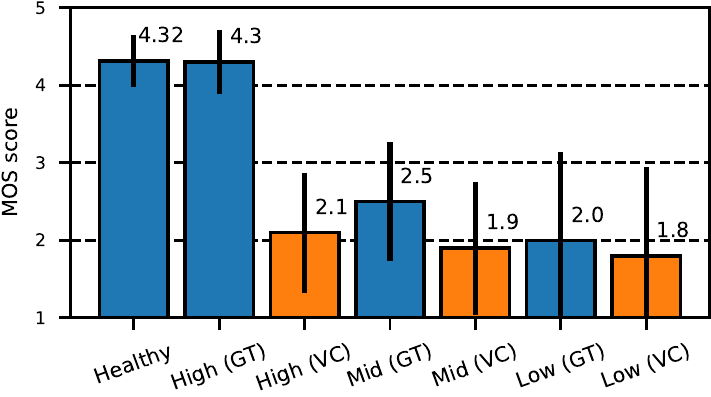}
    \caption{Mean opinion scores for naturalness grouped by intelligibility with 95\% confidence intervals. Blue denotes original, while orange denotes VC samples.}
    \label{fig:naturalness}
\end{figure}

The results of the naturalness experiments  are  presented in Figure \ref{fig:naturalness}, which shows the MOS score for each of the seven types of speech tested, grouped by intelligibility, and with their 95\% confidence intervals indicated. For clarity, the actual MOS scores are indicated on top of each bar. 

We first focus on the question how GT pathological speech affects the naturalness perceived by listeners which is measured by the MOS score (our RQ2). Figure \ref{fig:naturalness} shows that healthy speech and GT high intelligibility dysarthric speech have a similar MOS score. However, as intelligibility decreases, so does the MOS score, indicating that the MOS score not only captures naturalness but is influenced by the intelligibility of the speech. These results show that naive listeners cannot separate severity of a pathology and unnaturalness when asked to judge the naturalness of a speech sample. This also means that the GT MOS results are an upper bound on the achievable naturalness of synthetic pathological samples. 

Regarding the synthetic pathological speech, the performance on the high (VC) samples is somewhat lower than the performance of the HL-VQ-VAE-3 model on the VCC2020 challenge and identical to the performance of autoencoder-based models (2.1) \cite{ho2020non}. However, the type of stimuli is different, so the differences in MOS are not directly comparable. The difference is most likely due to channel differences, the decreased intelligibility of the speech, and the different sampling frequency (UASpeech is 16 kHz, while VCC2020 is 24 kHz). When we compare the MOS scores for the converted speech of the different intelligibility speakers, we observe a slight degradation in naturalness with decreasing intelligibility. Comparing the VC and GT results, however, we observe a large degradation for the converted high intelligibility speech (Wilcoxon signed-rank test: p $\leq$ $0.05$). The difference in VC and GT MOS scores for the mid and low intelligibility speakers is much smaller (Wilcoxon signed-rank test: mid p $\leq 0.05$, low p $\geq$ $0.05$). It is possible that the standard 5-point MOS does not allow to express the nuances between mid and low samples appropriately. Therefore, for future studies concerning naturalness of pathological speech, we would recommend using a slightly wider, 7-point scale. Returning to RQ1, we can conclude that the synthetic speech of mid and low intelligibility pathological speakers have a naturalness that is perceived similar to that of real pathological speech, while synthetic high intelligibility pathological speech is not perceived as being as natural as real high intelligibility pathological speech. 

To summarise, pathological speech is not perceived natural according to the MOS scale by naive listeners. In the case of mid and low intelligibility pathological speech, the perceived naturalness is similar between that of synthetic and real pathological speech. This is, however, not the case for high intelligibility synthesised pathological speech which is rated as being far less natural than real pathological speech.  The performance of the VC approach is comparable to the one observed with typical speakers, therefore the current method is accessible to typical speakers, however this does not mean that VC is accessible to typical speakers (see Section \ref{subsec:accessibility}).

\begin{figure*}[ht]
    \centering
    \includegraphics{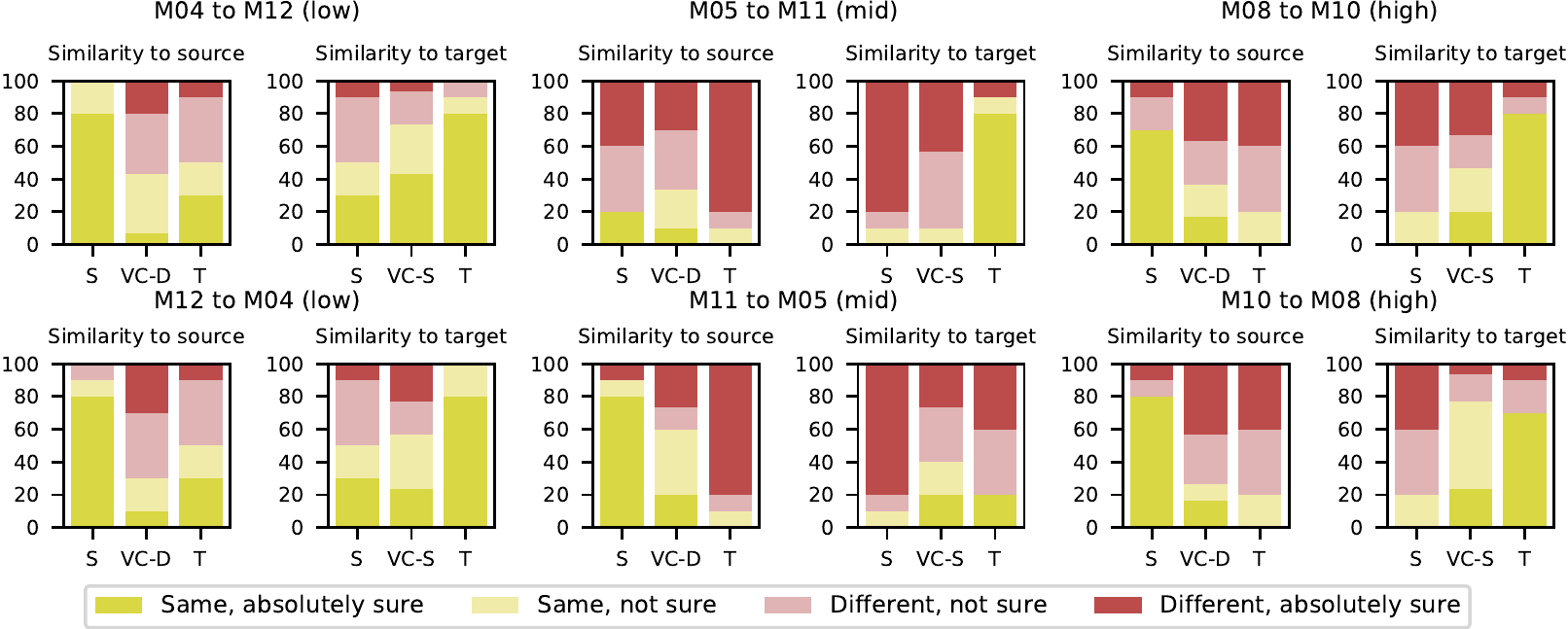}
    \caption{Results of the speaker similarity experiments grouped by intelligibility pairs. S stands for source, T for target, VC-D for voice conversion different (VC samples should be different from source) and VC-S for voice conversion same (VC samples should be same as target).}
    \label{fig:similarity}
\end{figure*}

\subsection{Similarity}
\label{sec:similarity_result}

This section presents and discusses the results of the similarity experiments in order to answer the question whether it is possible to convert voice characteristics of pathological speakers. The results are presented in Figure \ref{fig:similarity}. In each of the 12 panels, we visualise the results of comparing a voice converted (VC-D / VC-S) sample with the GT source (S) (Similarity to source) or the GT target (Similarity to target). Also, the GT samples are compared between them: S samples are compared to S samples to know how recognisable the source speaker is, T samples are compared to T samples to know how recognisable the target speaker is and S samples are compared to T samples in order to know how distinguishable is the source from the target speaker. Note that for each speaker pair in the top panel the source speaker is the target one in the bottom panel and vice versa, so this information appears repeated in Figure \ref{fig:similarity}. Additionally letters in the case of the VC comparisons are used to help interpretation of the figures: VC-D stands for VC-different (i.e when converting M04 to M12, the converted should be different from M04), VC-S stands for VC-same (similarly, when converting M04 to M12, the converted should be same as M12). 

For the low intelligibility pair (left 2 columns of Figure 3), the speakers seem reasonably distinguishable when looking at the GT as there is a 100\% of agreement that M04 samples are produced by M04 and 90\% for M12. For the speech samples of speaker M04 converted to speaker M12 (top panels), 73.33\% of the converted samples were indicated as being from speaker M12 (VC-S), meaning that the conversion is fairly successful for that pair. For the speech samples of speaker M12 converted to speaker M04 (bottom panels), 56.33\% of the converted M12-M04 samples (VC-S) were indicated as being from speaker M04. The results show that for the M12-M04 conversion the model is able to remove some of the source speaker (M12) characteristics and add some of the target (M04) ones, although to a lesser extent than in the M04-M12 conversion. Therefore, we conclude that the voice characteristic conversions for the low intelligibility speakers are successful.

For the mid intelligibility pair (middle four panels), the M11 seems to be clearly recognisable as there is a 90\% of agreement that M11 samples are produced by M11, however listeners have difficulties recognising the voice characteristics of M05, i.e., only 20\% of the trials where both samples were from speaker M05 were judged as both being from M05. For M05-M11 the VC performs poorly, which is indicated by 90\% perceiving it different from the target (VC-S result). For M05-M11 the VC-S reaches a 20\% of absolutely sure agreement. Notice that although it is a low score, it is the same that the GT samples exhibit. The voice characteristic conversions for the mid intelligibility speakers are thus inconclusive: while in one case the VC fails, in the other participants fail to recognise the speaker even from the GT samples. Further experimentation with more speaker pairs is needed.

For the high intelligibility pairs (right 2 columns of Figure 3), the speakers seem reasonably distinguishable. We can see that there is a 70\% of agreement that M08 samples are produced by M08 and an 80\% for M10. For M08-M10, there is a 46.66\% of agreement that the converted samples sound like M10. For M10 to M08 VC, 75\% of the listeners indicate that the converted samples sound like M08. We can see that some of the voice characteristics are successfully transferred for the high intelligibility samples, however while on the conversions M10 to M08 the result is similar to the GT samples, on the other direction (M08 to M10) there is a gap of 33.33\% with respect to the GT. This behaviour is the same that we observed with low intelligibility pair conversions: although the speakers from the same pair are recognised with a similar agreement (100\% and 90\% for low intelligibility and 80\% and 70\% for the high intelligibility) the conversions are more successful in one direction than on the other.

\subsection{Limitations of the proposed approach}
\label{subsec:limitations}

An assumption of the proposed approach is that the speaker identity is not affected by the speech pathology, which is certainly untrue for speech pathologies which are dysphonic, i.e. where the voice characteristics are known to be affected. By performing AB testing with GT speakers, we have tried to account for these scenarios in the perceptual evaluations. From the speaker similarity experiment, we have seen that in some cases (i.e., M05) listeners had difficulties of recognising the voice characteristics even in the GT. These results confirm that the proposed approach cannot be used for all types of speech pathologies. To solve this issue, we would need to have a deeper understanding of what happens to the speaker characteristics in these speech pathologies. For example, the speaker embeddings themselves could be used to predict the new pathological speaker embeddings of the same speaker, transformed according to the vocal pathology (i.e. type of dysphonia).

\subsection{Accessibility of VC to atypical speakers}
\label{subsec:accessibility}
VC of atypical speech produced similar naturalness in the high intelligibility case as typical speech on VQ-VAE based methods. Nevertheless, we see that there is room for improvement compared to typical speech, as other studies employing certain non-parallel VC approaches can achieve human-like naturalness. Unfortunately, these VC approaches cannot easily be used for our task as they often leverage linguistic features or ASR bottleneck features \cite{Liu2018, tian2018average}. The need for ASR features is especially problematic as these features are extracted from ASR systems, whose performance on atypical speech is generally much worse than that on typical speech, meaning that the quality of these extracted features are also expected to be lower for these speakers. Therefore, we conclude that accessibility to VC is limited for atypical speakers, but this is because parallel and ASR-based techniques can hardly be used by them.    

\section{Conclusions}

In this paper, we propose a new approach to pathological speech synthesis, by customising an existing pathological speech sample to a new speaker's voice characteristics. In order to do this pathological-to-pathological speech conversion, we use an autoencoder-based voice conversion (VC) technique. 
When comparing our results with the ones obtained in the VCC2020 challenge dataset \cite{ho2020non}, we can see that ours are somewhat lower, which is most likely due to channel differences, the decrease in the speech intelligibility and the different sampling rate.  
We find that even real pathological speech seems to affect perceived naturalness as shown by MOS scores, meaning that there is a bound on achievable naturalness for pathological speech conversion. 
Overall, we observe a decreasing trend in MOS with decreasing intelligibility. Therefore, for low and mid intelligibility, the difference in perceived naturalness between real and VC is small.
Conversion of voice characteristics for low intelligibility speakers is successful, for high intelligibility it is also possible to transfer the voice characteristics partially. However, more experimentation is needed for the mid intelligibility with more speakers: we experienced that in one case the VC failed, and on the other participants fail to recognise the speaker even from the real recordings. Whether    the differences in the results for the different intelligibility levels is due to the intelligibility levels or due to other speech characteristics needs to be further investigated. The question of pathological intergender (male to female) and female VC also needs to be investigated. The performance of the approach is comparable to the one observed with typical speakers, therefore the current method is accessible to atypical speakers. However, in the paper, we outlined some issues such as the need for linguistic resources and parallel data, as an obstacle for more natural VC for pathological speakers.

\section{Acknowledgements}

B.M.H. is funded through the EU’s H2020 research and innovation programme under MSC grant agreement No 766287. The Department of Head and Neck Oncology and Surgery of the Netherlands Cancer
Institute receives a research grant from Atos Medical (H\"orby, Sweden), which contributes to the existing infrastructure for quality of life research.

\bibliographystyle{IEEEtran}

%\newpage
\bibliography{mybib}

% Generated by IEEEtran.bst, version: 1.13 (2008/09/30)
\begin{thebibliography}{10}
\providecommand{\url}[1]{#1}
\csname url@samestyle\endcsname
\providecommand{\newblock}{\relax}
\providecommand{\bibinfo}[2]{#2}
\providecommand{\BIBentrySTDinterwordspacing}{\spaceskip=0pt\relax}
\providecommand{\BIBentryALTinterwordstretchfactor}{4}
\providecommand{\BIBentryALTinterwordspacing}{\spaceskip=\fontdimen2\font plus
\BIBentryALTinterwordstretchfactor\fontdimen3\font minus
  \fontdimen4\font\relax}
\providecommand{\BIBforeignlanguage}[2]{{%
\expandafter\ifx\csname l@#1\endcsname\relax
\typeout{** WARNING: IEEEtran.bst: No hyphenation pattern has been}%
\typeout{** loaded for the language `#1'. Using the pattern for}%
\typeout{** the default language instead.}%
\else
\language=\csname l@#1\endcsname
\fi
#2}}
\providecommand{\BIBdecl}{\relax}
\BIBdecl

\bibitem{vachhani2018data}
B.~Vachhani, C.~Bhat, and S.~K. Kopparapu, ``Data augmentation using healthy
  speech for dysarthric speech recognition.'' in \emph{Interspeech}, 2018, pp.
  471--475.

\bibitem{harvillsynthesis}
J.~Harvill, D.~Issa, M.~Hasegawa-Johnson, and C.~Yoo, ``Synthesis of new words
  for improved dysarthric speech recognition on an expanded vocabulary,'' in
  \emph{International Conference on Acoustics, Speech and Signal Processing},
  2021.

\bibitem{jiao2018simulating}
Y.~Jiao, M.~Tu, V.~Berisha, and J.~Liss, ``Simulating dysarthric speech for
  training data augmentation in clinical speech applications,'' in \emph{2018
  IEEE international conference on acoustics, speech and signal processing
  (ICASSP)}.\hskip 1em plus 0.5em minus 0.4em\relax IEEE, 2018, pp. 6009--6013.

\bibitem{halpern2021evaluation}
B.~M. Halpern, J.~Fritsch, E.~Hermann, R.~van Son, O.~Scharenborg, and M.-M.
  Doss, ``An objective evaluation framework for pathological speech
  synthesis,'' \emph{Submitted to Signal Processing Letters}, 2021.

\bibitem{Halpern2020}
\BIBentryALTinterwordspacing
B.~M. Halpern, R.~van Son, M.~van~den Brekel, and O.~Scharenborg, ``{Detecting
  and Analysing Spontaneous Oral Cancer Speech in the Wild},'' in \emph{Proc.
  Interspeech 2020}, 2020, pp. 4826--4830. [Online]. Available:
  \url{http://dx.doi.org/10.21437/Interspeech.2020-1598}
\BIBentrySTDinterwordspacing

\bibitem{rudzicz2013adjusting}
F.~Rudzicz, ``Adjusting dysarthric speech signals to be more intelligible,''
  \emph{Computer Speech \& Language}, vol.~27, no.~6, pp. 1163--1177, 2013.

\bibitem{aryal2016data}
S.~Aryal and R.~Gutierrez-Osuna, ``Data driven articulatory synthesis with deep
  neural networks,'' \emph{Computer Speech \& Language}, vol.~36, pp. 260--273,
  2016.

\bibitem{kim2008dysarthric}
H.~Kim, M.~Hasegawa-Johnson, A.~Perlman, J.~Gunderson, T.~S. Huang, K.~Watkin,
  and S.~Frame, ``Dysarthric speech database for universal access research,''
  in \emph{Ninth Annual Conference of the International Speech Communication
  Association}, 2008.

\bibitem{rudzicz2012torgo}
F.~Rudzicz, A.~K. Namasivayam, and T.~Wolff, ``{The TORGO database of acoustic
  and articulatory speech from speakers with dysarthria},'' \emph{Language
  Resources and Evaluation}, vol.~46, no.~4, pp. 523--541, 2012.

\bibitem{middag2012automatic}
C.~Middag, ``Automatic analysis of pathological speech,'' Ph.D. dissertation,
  Ghent University, 2012.

\bibitem{Hsu2017}
\BIBentryALTinterwordspacing
W.-N. Hsu, Y.~Zhang, and J.~Glass, ``Learning latent representations for speech
  generation and transformation,'' in \emph{Proc. Interspeech 2017}, 2017, pp.
  1273--1277. [Online]. Available:
  \url{http://dx.doi.org/10.21437/Interspeech.2017-349}
\BIBentrySTDinterwordspacing

\bibitem{kaneko2018cyclegan}
T.~Kaneko and H.~Kameoka, ``Cyclegan-vc: Non-parallel voice conversion using
  cycle-consistent adversarial networks,'' in \emph{2018 26th European Signal
  Processing Conference (EUSIPCO)}.\hskip 1em plus 0.5em minus 0.4em\relax
  IEEE, 2018, pp. 2100--2104.

\bibitem{kaneko2019cyclegan}
T.~Kaneko, H.~Kameoka, K.~Tanaka, and N.~Hojo, ``{CycleGAN-VC2: Improved
  CycleGAN-based non-parallel voice conversion},'' in \emph{ICASSP 2019-2019
  IEEE International Conference on Acoustics, Speech and Signal Processing
  (ICASSP)}.\hskip 1em plus 0.5em minus 0.4em\relax IEEE, 2019, pp. 6820--6824.

\bibitem{kaneko2020cyclegan}
\BIBentryALTinterwordspacing
{Kaneko, Takuhiro and Kameoka, Hirokazu and Tanaka, Kou and Hojo, Nobukatsu},
  ``{CycleGAN-VC3: Examining and Improving CycleGAN-VCs for Mel-Spectrogram
  Conversion},'' in \emph{Proc. Interspeech 2020}, 2020, pp. 2017--2021.
  [Online]. Available: \url{http://dx.doi.org/10.21437/Interspeech.2020-2280}
\BIBentrySTDinterwordspacing

\bibitem{ho2020non}
T.~V. Ho and M.~Akagi, ``Non-parallel voice conversion based on hierarchical
  latent embedding vector quantized variational autoencoder,'' in \emph{Proc.
  Joint Workshop for the Blizzard Challenge and Voice Conversion Challenge
  2020}, 2020, pp. 140--144.

\bibitem{feng2021quantifying}
S.~Feng, O.~Kudina, B.~M. Halpern, and O.~Scharenborg, ``Quantifying bias in
  automatic speech recognition,'' \emph{arXiv preprint arXiv:2103.15122}, 2021.

\bibitem{adda2005speech}
M.~Adda-Decker and L.~Lamel, ``{Do speech recognizers prefer female
  speakers?}'' in \emph{Ninth European Conference on Speech Communication and
  Technology}, 2005.

\bibitem{moro2019study}
\BIBentryALTinterwordspacing
L.~Moro-Velazquez, J.~Cho, S.~Watanabe, M.~A. Hasegawa-Johnson, O.~Scharenborg,
  H.~Kim, and N.~Dehak, ``{Study of the Performance of Automatic Speech
  Recognition Systems in Speakers with Parkinson’s Disease},'' in \emph{Proc.
  Interspeech 2019}, 2019, pp. 3875--3879. [Online]. Available:
  \url{http://dx.doi.org/10.21437/Interspeech.2019-2993}
\BIBentrySTDinterwordspacing

\bibitem{koenecke2020racial}
A.~Koenecke, A.~Nam, E.~Lake, J.~Nudell, M.~Quartey, Z.~Mengesha, C.~Toups,
  J.~R. Rickford, D.~Jurafsky, and S.~Goel, ``Racial disparities in automated
  speech recognition,'' \emph{Proceedings of the National Academy of Sciences},
  vol. 117, no.~14, pp. 7684--7689, 2020.

\bibitem{hermann2020dysarthric}
E.~Hermann and M.~M. Doss, ``{Dysarthric speech recognition with lattice-free
  MMI},'' in \emph{ICASSP 2020-2020 IEEE International Conference on Acoustics,
  Speech and Signal Processing (ICASSP)}.\hskip 1em plus 0.5em minus
  0.4em\relax IEEE, 2020, pp. 6109--6113.

\bibitem{veaux2017cstr}
C.~Veaux, J.~Yamagishi, K.~MacDonald \emph{et~al.}, ``{CSTR VCTK Corpus:
  English Multi-speaker Corpus for CSTR Voice Cloning Toolkit (2017)},''
  \emph{URL http://dx. doi. org/10.7488/ds}, 2017.

\bibitem{mcfee2015librosa}
B.~McFee, C.~Raffel, D.~Liang, D.~P. Ellis, M.~McVicar, E.~Battenberg, and
  O.~Nieto, ``librosa: Audio and music signal analysis in python,'' 2015.

\bibitem{tim_sainburg_2019_3243139}
\BIBentryALTinterwordspacing
T.~Sainburg, ``timsainb/noisereduce: v1.0,'' Jun. 2019. [Online]. Available:
  \url{https://doi.org/10.5281/zenodo.3243139}
\BIBentrySTDinterwordspacing

\bibitem{skerry2018towards}
R.~Skerry-Ryan, E.~Battenberg, Y.~Xiao, Y.~Wang, D.~Stanton, J.~Shor, R.~Weiss,
  R.~Clark, and R.~A. Saurous, ``{Towards end-to-end prosody transfer for
  expressive speech synthesis with Tacotron},'' in \emph{international
  conference on machine learning}.\hskip 1em plus 0.5em minus 0.4em\relax PMLR,
  2018, pp. 4693--4702.

\bibitem{yamamoto2020parallel}
R.~Yamamoto, E.~Song, and J.-M. Kim, ``{Parallel WaveGAN: A fast waveform
  generation model based on generative adversarial networks with
  multi-resolution spectrogram},'' in \emph{ICASSP 2020-2020 IEEE International
  Conference on Acoustics, Speech and Signal Processing (ICASSP)}.\hskip 1em
  plus 0.5em minus 0.4em\relax IEEE, 2020, pp. 6199--6203.

\bibitem{yi2020voice}
Z.~Yi, W.-C. Huang, X.~Tian, J.~Yamagishi, R.~K. Das, T.~Kinnunen, Z.~Ling, and
  T.~Toda, ``Voice conversion challenge 2020—intra-lingual semi-parallel and
  cross-lingual voice conversion—,'' in \emph{Proc. Joint Workshop for the
  Blizzard Challenge and Voice Conversion Challenge 2020}, 2020, pp. 80--98.

\bibitem{toda2016voice}
T.~Toda, L.-H. Chen, D.~Saito, F.~Villavicencio, M.~Wester, Z.~Wu, and
  J.~Yamagishi, ``The voice conversion challenge 2016.'' in \emph{Interspeech},
  2016, pp. 1632--1636.

\bibitem{Liu2018}
\BIBentryALTinterwordspacing
L.-J. Liu, Z.-H. Ling, Y.~Jiang, M.~Zhou, and L.-R. Dai, ``{WaveNet Vocoder
  with Limited Training Data for Voice Conversion},'' in \emph{Proc.
  Interspeech 2018}, 2018, pp. 1983--1987. [Online]. Available:
  \url{http://dx.doi.org/10.21437/Interspeech.2018-1190}
\BIBentrySTDinterwordspacing

\bibitem{tian2018average}
X.~Tian, J.~Wang, H.~Xu, E.~S. Chng, and H.~Li, ``{Average Modeling Approach to
  Voice Conversion with Non-Parallel Data.}'' in \emph{Odyssey}, vol. 2018,
  2018, pp. 227--232.

\end{thebibliography}

\end{document}